\documentclass[%
 reprint,
superscriptaddress,
amsmath,amssymb,
aps,
prl,
]{revtex4-2}

\usepackage{graphicx}
\usepackage{dcolumn}
\usepackage{bm}
\usepackage{color}
\usepackage[normalem]{ulem}

\usepackage{hyperref}

\def\mueff{$\mu_{eff}$=2.3(7)~$\mu_B$}
\def\fwhm{$\Delta E_{FWHM}$=3.0(2) meV}

\begin{document}

\preprint{APS/123-QED}

\title{Connection between $f$-electron correlations and magnetic excitations in UTe$_2$}
\author{Thomas Halloran}
\affiliation{NIST Center for Neutron Research, Gaithersburg, Maryland\ 20899, USA}
\affiliation{Department of Physics and Astronomy, University of Maryland}
\author{Peter Czajka}
\affiliation{NIST Center for Neutron Research, Gaithersburg, Maryland\ 20899, USA}
\affiliation{Department of Physics and Astronomy, University of Maryland}
\author{Gicela Saucedo Salas}
\affiliation{NIST Center for Neutron Research, Gaithersburg, Maryland\ 20899, USA}
\affiliation{Department of Physics and Astronomy, University of Maryland}
\author{Corey Frank}
\affiliation{NIST Center for Neutron Research, Gaithersburg, Maryland\ 20899, USA}
\affiliation{Department of Physics and Astronomy, University of Maryland}
\author{Chang-Jong Kang}
\affiliation{Department of Physics, Chungnam National University, Daejeon 34134, Korea}
\author{J. A. Rodriguez-Rivera}
\affiliation{NIST Center for Neutron Research, Gaithersburg, Maryland\ 20899, USA}
\affiliation{Department of Materials Science, University of Maryland, College Park, MD 20742}
\author{Jakob Lass}
\affiliation{Laboratory for Neutron Scattering and Imaging, Paul Scherrer Institut, CH-5232 Villigen PSI, Switzerland}
\affiliation{Laboratory for Quantum Magnetism, Institute of Physics, Ecole Polytechnique Fédérale de Lausanne (EPFL), CH-1015 Lausanne, Switzerland}
\author{Daniel G. Mazzone}
\affiliation{Laboratory for Neutron Scattering and Imaging, Paul Scherrer Institut, CH-5232 Villigen PSI, Switzerland}
\author{Marc Janoschek}
\affiliation{ Laboratory for Neutron and Muon Instrumentation, PSI Center for Neutron and Muon Sciences, Paul Scherrer Institute, 5232 Villigen-PSI}
\affiliation{Physik-Institut, Universität Zürich, Winterthurerstrasse 190, CH-8057 Zürich, Switzerland}
\author{Gabi Kotliar}
\affiliation{Department of Physics and Astronomy, Rutgers University, Piscataway, New Jersey 08854, USA}
\affiliation{Condensed Matter Physics and Materials Science Department, Brookhaven National Laboratory, Upton, New York 11973, USA}
\author{Nicholas P. Butch}
\affiliation{NIST Center for Neutron Research, Gaithersburg, Maryland\ 20899, USA}
\affiliation{Department of Physics and Astronomy, University of Maryland}

\date{\today}

\begin{abstract}
The detailed anisotropy of the low-temperature, low-energy magnetic excitations of the candidate spin-triplet superconductor UTe$_2$ is revealed using inelastic neutron scattering. The magnetic excitations emerge from the Brillouin zone boundary at the high symmetry $Y$ and $T$ points and disperse along the crystallographic $\hat{b}$-axis. In applied magnetic fields to at least $\mu_0 H=11$~T along the $\hat{c}-$axis, the magnetism is found to be field-independent in the $(hk0)$ plane. The scattering intensity is consistent with that expected from U$^{3+}$/U$^{4+}$ $f$-electron spins with preferential orientation along the crystallographic $\hat{a}$-axis, and a fluctuating magnetic moment of \mueff. These characteristics indicate that the excitations are due to intraband spin excitons arising from $f$-electron hybridization.

\end{abstract}

\maketitle
\section{Introduction}
The heavy-fermion paramagnet UTe$_2$ has been a subject of focused investigation within the strongly correlated electron community since the recent discovery of its highly nontrivial superconducting ground state~\cite{ran_nearly_2019}. UTe$_2$ strongly violates the Pauli limit for BCS superconductors, which relates the temperature of the onset of superconductivity $T_c\approx2~K$~\cite{sakai_single_2022,aoki_molten_2024} to the critical magnetic field. The Pauli limit is exceeded by factors of approximately 2, 4, and 2.5 for fields along the crystalline $\hat{a}$, $\hat{b}$, and $\hat{c}$ directions respectively, and magnetization scaling strongly suggests proximity to ferromagnetism~\cite{ran_nearly_2019,tokiwa_reinforcement_2024}, which was taken as initial supporting evidence of a spin-triplet superconducting ground state in UTe$_2$. 

Nuclear magnetic resonance (NMR) studies indicate that the spin susceptibility remains constant upon cooling below $T_c$~\cite{fujibayashi_superconducting_2022,nakamine_anisotropic_2021,ambika_possible_2022}. This violates expectations for spin-singlet BCS superconductivity, where a Knight shift signifies is associated with the condensation of the superconducting quasiparticles. This, along with other experimental results such as polar Kerr effect measurements\cite{hayes_multicomponent_2021,wei_interplay_2022,Ajeesh_Rosa_Thomas_2023}, exotic reentrant superconductivity at fields above 35 T~\cite{ran_extreme_2019,lewin_high-field_2024},  and fluctuations in $\mu$SR~\cite{sundar_coexistence_2019,azari_absence_2023,sundar_ubiquitous_2023}, are all taken to be supporting evidence for a potential spin-triplet superconducting ground state in UTe$_2$. Such a superconducting pairing is exceedingly rare in condensed matter systems, with the only conclusive example being that of the rigorously studied superfluid phase of $^3$He~\cite{leggett_theoretical_1975}. A spin-triplet superconductor is of particular interest in the context of applications to quantum devices. 

Many open questions remain regarding the ground state superconductivity in UTe$_2$. The superconducting nodal gap function~\cite{suetsugu_fully_2024,ishizuka_insulator-metal_2019,ishihara_chiral_2023}, the order parameters associated with each of the superconducting phases~\cite{hayes_multicomponent_2021,fujibayashi_superconducting_2022}, time reversal symmetry breaking~\cite{wei_interplay_2022}, the role of magnetic interactions, and the electronic band structure and Fermi surface are all unresolved. As spin-triplet superconductivity can be mediated by ferromagnetism (although an antiferromagnetic mechanism~\cite{hu_spin-singlet_2021,Kreisel_Quan_Hirschfeld_2022} is also possible), unambiguous understanding of the spin interactions is required to describe the ground state. Additionally, band hybridization in heavy fermion systems involves the Kondo effect and RKKY interactions, and the delicate interplay of these effects with $f$-electron band has been linked to unconventional superconductivity and conduction electron mediated magnetic exchange interactions~\cite{White_Thompson_Maple_2015}. Their role in UTe$_2$ is currently not well understood. 

Inelastic neutron scattering (INS) experiments are the most direct probe of spin fluctuations and magnetism in heavy fermion superconductors, many of which possess many similarities to UTe$_2$. In the cases of UCoGe~\cite{stock_anisotropic_2011,prokes_anomalous_2010} and UGe$_2$~\cite{haslbeck_ultrahigh-resolution_2019,prokes_anomalous_2010}, ferromagnetic fluctuations are observed, whereas antiferromagnetic correlations associated with superconductivity have been observed in materials like URu$_2$Si$_2$~\cite{butch_symmetry_2015,broholm_magnetic_1991} and the  cuprates~\cite{fujita_progress_2012}. However, the imaginary susceptibility measured by INS $\chi''(\bm{Q},\hbar\omega)$ in correlated $f$-electron systems often arises not only from magnetic excitations like magnons, but from band excitons as in the cases of CePd$_3$~\cite{goremychkin_coherent_2018} and SmB$_6$\cite{fuhrman_interaction_2015}. Further, it has recently been demonstrated that in $f$-electron systems even the magnons in ordered phases sensitively depend upon the underlying correlated band structure~\cite{Simeth_Janoschck_2023}. An applied magnetic field can help to distinguish between magnons and band excitons. For example, in the case of CeRu$_2$Si$_2$~\cite{rossat-mignod_inelastic_1988}, magnetic field strongly suppresses magnon scattering due to interionic correlations, as opposed to the gentle perturbations expected in the band exciton picture. 

Earlier INS studies of UTe$_2$ report a condensation of excitations emerging at the Brillouin zone boundary upon cooling below the Kondo hybridization temperature of $T_K\approx40~K$ \cite{butch_symmetry_2022}. Additional INS results consistently report the presence of a broad continuum of excitations in the superconducting state peaked in intensity at energies of $\hbar\omega=4$ meV~\cite{knafo_low-dimensional_2021,duan_resonance_2021}, where $\hbar\omega$ denotes neutron energy transfer, with the scattering being confined to the $(hk0)$ plane. Additionally, an excitation observed a $\hbar\omega$=1 meV was suggested to be a resonant excitation which originates from a bound state within the particle-hole continuum gap~\cite{duan_resonance_2021,raymond_feedback_2021}. This was taken as evidence that the spin fluctuations were primarily antiferromagnetic in nature, as one would expect ferromagnetic fluctuations at integer $\bm{Q}$ vectors, where $\bm{Q}$ denotes momentum transfer.

Here, we report INS results on UTe$_2$ in the $(hk0)$ scattering plane with an applied field along the $\hat{c}$ axis. Our results resolve the low energy excitations across multiple Brillouin zones in the $(hk0)$ scattering plane at energies below 6 meV, accounting for nearly all the expected spectral weight from the magnetic moment of U$^{3+}$. The measured scattering intensity is independent to an applied magnetic field of $\mu_0 H\leq$11 T, suggesting that the magnetic response is dominated by interband excitations excited across the $df$ Kondo hybridization gap, rather than ferromagnetic or antiferromagnetic exchange between local spins.

\begin{figure}[t]
    \centering
    \includegraphics[width=1.0\columnwidth ]{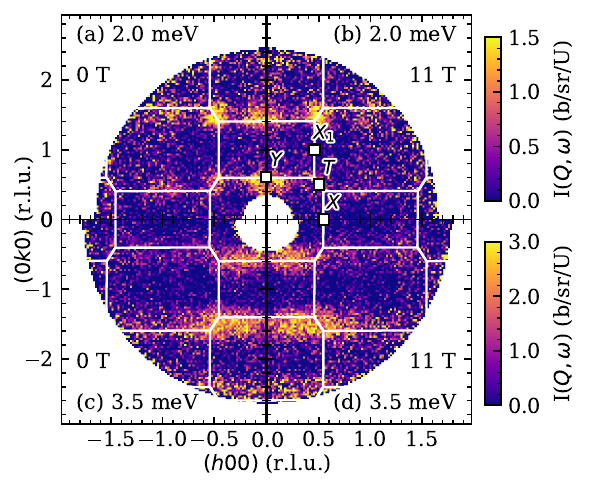}
    \caption{Symmetrized constant energy slices of inelastic neutron scattering of UTe2 at a nominal temperature of T=45~mK. The energy integration windows are $\hbar\omega\in[1.7,2.3]$ meV in (a,b), and $\hbar\omega\in[3.2,3.8]$ meV in (c,d). All observable scattering at the lowest energies originates from the $Y$ $(0\frac{1}{2}0)$ and $T$ ($\frac{1}{2}\frac{1}{2}$0) points, as shown in (a,b), with the lowest energy scattering emerging from the points denoted as Y (edge) and T (corner) type points. The magnetic field in (b,d) is applied along the $\hat{c}$-axis.}
    \label{fig:constE}
\end{figure}

\begin{figure}[t]
    \centering
    \includegraphics[]{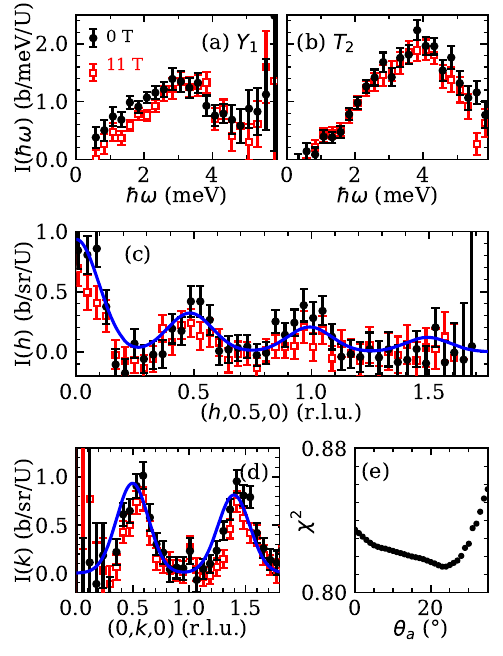}
    \caption{(a,b) Constant Q cuts along the energy dimension of UTe2 scattering integrated in a circular window of diameter $\Delta Q=0.2~\AA^{-1}$, centered at the Y1 (0,0.6,0) and T2 (0.4,1.4,0) points. Black filled circles represent 0 T measurements, red unfilled square makers 11 T. (c) Intensity integrated in the low-energy portion of the measurement of 1 meV to 2 meV. The cut is along the BZ edge, with an integration with in Q of 0.3 $\AA^{-1}$. (d) Scattering intensity in the same energy window along the $(0k0)$ direction. Both cuts are fit simultaneously using a form described in the text, with the result shown in blue. (e) Goodness of fit $\chi^2$ value for the cuts in (c) and (d) as a function of the magnetic moment direction, where $\theta_a$ is the tilt away from the \textit{a} axis towards the \textit{b} axis. All error bars represent one standard deviation.}
    \label{fig:cutfig}
\end{figure}

Scattering integrated over the elastic window ($\hbar\omega\in$[-0.15,0.15] meV) shows no evidence of magnetic or field-dependent scattering. A mosaic spread of full width at half maximum (FWHM) 5.0(1)$^o$ in the $(hk0)$ plane is determined from the nuclear Bragg peaks. A sample environment background is determined using a momentum averaged scattering from the regions in $\bm{Q},\omega$ space where magnetic scattering from the sample is taken to be absent, which is described in detail in the Supplementary Information (SI). 

Two representative constant energy slices of the magnetic scattering from UTe$_2$ are shown in Fig.~\ref{fig:constE}. The scattering in all cases has been symmetrized by the two-fold C$_{2m}$ lattice symmetry to enhance statistics. The lowest energy contribution to the scattering is shown in Fig.~\ref{fig:constE}(a,b). As in previous studies, the scattering is constrained to the Brillouin Zone (BZ) edge, emanating from the corners ($T$=($\frac{1}{2},\frac{1}{2}$,0)-like points) and edges ($Y$=($0,0.6,0$)-type points). Scattering at higher Brillioun zones is also observed.

\begin{figure*}[t]
    \centering
    \includegraphics[width=1.0\textwidth]{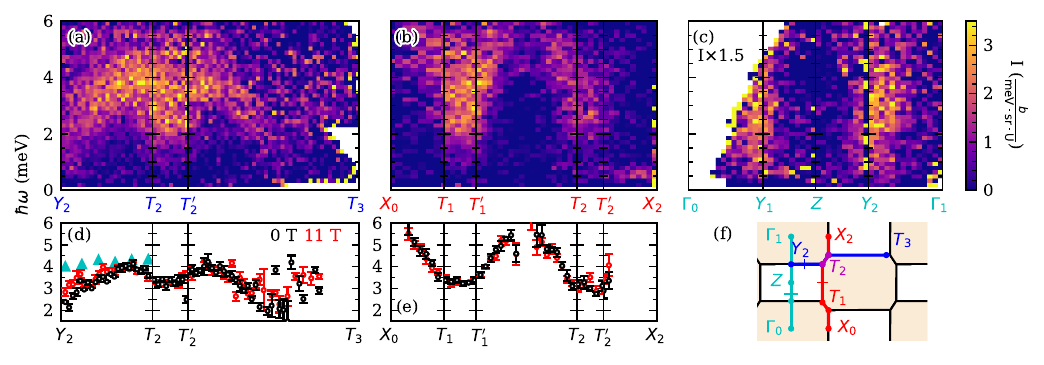}
    \caption{(a,b) Dispersion of magnetic excitation in UTe$_2$ using paths along the BZ edge as depicted in (f). The path in Q for panel (a) is primarily along the (h,1.5,0) direction and is in blue, and the path for (b) is primarily along the (0.5,k,0) direction and is shown in red. At every point in Q, a Lorentzian form is used to fit the intensity maxima as a function of energy transfer, giving effective dispersions shown in (d) and (e). (d) corresponds to subplot (a), and (e) corresponds to (b). Values of Q with insufficient statistics to find a stable intensity maxima have been omitted from these plots. Cyan triangles in (d) are the dispersion observed in Ref.~\cite{butch_symmetry_2022}. (c) Magnetic excitation dispersion along the $(0,k,0)$ direction through the zone center, as depicted by the cyan path in (f). All error bars represent one standard deviation. }
    \label{fig:dispersion}
\end{figure*}

Constant energy slices in energy window of maximum scattering intensity $\hbar\omega\in\{3.2,3.8\}$ meV are shown for the 0 T measurement in Fig.~\ref{fig:constE}(c) and 11 T in Fig.~\ref{fig:constE}(d). Here the scattering remains on the BZ edges, but the excitations have dispersed such that they have smeared in the $(h00)$ direction. A minimum in intensity has emerged along the $(0k0)$ line and zero intensity is observed at all $\Gamma$ points. As is made more quantitatively clear in Fig.\ref{fig:cutfig}, there is no qualitative difference in scattering intensity for the 0 T and 11 T settings apart from a small deviation shown in Fig~\ref{fig:cutfig}(a). This also true for the elastic scattering, as shown in the SI. 

In Fig.~\ref{fig:cutfig}, representative intensity cuts along $\bm{Q}$ and energy transfer are shown. All data presented in cuts is not symmetrized. Fig.~\ref{fig:cutfig}(a) and (b) plot constant $\bm{Q}$ cuts along energy at the high symmetry points $Y_1=(0,0.6,0)$ in (a) and $T_2=(0.4,1.4,0)$ in (b). Both cuts are integrated in the $Q$ dimension in a circle of radius $\delta Q$=0.3$~\AA^{-1}$ ($h\pm$0.2 r.l.u., $k\pm$0.29 r.l.u.). The energy dependence of the intensity in both plots shows a monotonic increase in intensity from $\hbar\omega=0$, with a peak of intensity at 3.2(1) meV at $Y_1$ and 4.0(1) meV $T_2$ meV respectively. There is no statistically significant field-dependence to the scattering at the $T_2$ point. At energies below $\hbar\omega=3~$meV, the zero field scattering is slightly more intense at the $Y_1$ point, as shown in Fig.~\ref{fig:cutfig}(a). The integrated intensity within $\hbar\omega\in\{0.3,3.0\}$ is 2.0(3) b/U for $\mu_0 H=$0 T, and 1.5(3) b/U for $\mu_0 H=$11 T. We are unable to resolve the proposed spin-resonance excitation reported in Refs. \cite{duan_resonance_2021,knafo_low-dimensional_2021}. 

Fig.~\ref{fig:cutfig}(c) and (d) show constant energy cuts integrated in $\hbar\omega\in\{1,2\}$ meV. This integration window accounts for most of the scattering before dispersion becomes significant. Consistent with previous studies~\cite{duan_incommensurate_2020,duan_resonance_2021,butch_symmetry_2022,raymond_feedback_2021,knafo_low-dimensional_2021}, we find broad peaks in intensity centered at the Brillouin zone edges. Again, there is no difference between the 0 T and 11 T scattering intensities. In (c) the cut is along the $(h,1.5,0)$ direction, i.e. along the BZ edge, and in (d) the cut is along the $(0,k,0)$ direction. In both cases the scattering was integrated in a perpendicular $\bm{Q}$ range of 0.3$~\AA^{-1}$. The cuts are simultaneously fit to a model of nearest neighbor correlated spins described later in the text. 

Finally, Fig.~\ref{fig:dispersion}(a) and (b) present the full dispersion of the magnetic excitations in UTe$_2$ along the zone boundary, with Fig~\ref{fig:dispersion}(c) including the $\Gamma_0=(000)$, $\Gamma_1=(010)$, and $\Gamma_2=(020)$ points. The $\bm{Q}$ path in Fig.~\ref{fig:dispersion}(a) is primarily along the $(h00)$ direction and is depicted by the blue path in the schematic of the Brillouin zone in (c), and (b) is along the $(0k0)$ direction and depicted by the red path in (f). The third path along $(0k0)$ is shown in (c), revealing no  dispersion at the $Y$-type points and zero intensity at the zone center. The scattering is broader in energy than the expected instrumental resolution ($\delta E_{FWHM}=0.11$ meV for $\hbar\omega$=0~meV), and at every point in $\bm{Q}$ the energy dependent intensity $I(\hbar\omega)$ may be fit to the form of a simple Lorentzian peak of FWHM width \fwhm. For all values of $\bm{Q}$ with significant enough intensity to perform this fit, the extracted dispersion of magnetic excitations along the representative paths in $\bm{Q}$ is presented in Figs.~\ref{fig:dispersion}(d) and (e). Along both directions, the peak in scattering intensity has dispersion with clear periodicity. Due to the magnetic form factor of U$^{3+/4+}$ and a net spin polarization factor in the neutron scattering cross-section, these excitations become faint at the highest accessible values of $\bm{Q}$. The dispersion is steeper along the $(0k0)$ direction, as shown both in Figs.~\ref{fig:dispersion}(b) and (e), and in all cases peaks at the $T$ points.
\subsection{Analysis}
Overall, the scattering presented is compatible with previous INS studies in the superconducting phase. There are two significant differences. First, the resonant excitation reported in Refs. \cite{duan_resonance_2021} is not observable in our measurement. While this resonance was initially reported at the $Y_1=(0,0.6,0)$ point, Independent measurements also showed the excitation at the $Y_2=(0,1.4,0)$ point~\cite{raymond_feedback_2021}. Here, our results do not show evidence of a peak in scattering at $\hbar\omega$=1 meV at either $Y_1$ or $Y_2$. While a more detailed examination is shown in the SI, we conclude that no resonance is observed in our measurement.

Secondly, these data allow for the full extraction of the dispersion of magnetic excitations in the  $(hk0)$ scattering plane. An important result that we extract from this is the quantity of the fluctuating moment for both the 0 T and 11 T configurations. Because the scattering has been normalized, the well-known total moment sum rule~\cite{hohenberg_sum_1974,zaliznyak2005magnetic} of the form 
\begin{equation}
    \int S(\bm{Q},\omega)d\bm{Q}d\omega/\int d\bm{Q}=S(S+1)
\end{equation}.

The dynamical structure factor $S(\bm{Q},\omega)$ is related to the scattering intensity by 
\begin{equation}
    I(\bm{Q},\omega) = \frac{k_f}{k_i}|\frac{g}{2} F(\bm{Q})|^2 r_0^2 \sum_{\alpha,\beta}(\delta_{\alpha\beta} - \frac{Q_\alpha Q_\beta}{Q^2}) S^{\alpha\beta}(\bm{Q},\omega).
\end{equation}
Here, the scattering has been normalized by the kinematic ratio of final and initial momenta $\frac{k_f}{k_i}$, the Lande g-factor $g$ is taken to be 2, $|F(Q)|^2$ is the modulus of the magnetic form factor associated with scattering from U$^{3+}$ or U$^{4+}$, which are indistinguishable in this measurement, $r_0=5.391\cdot10^{-13}$ cm is the characteristic magnetic neutron scattering length, and the summation is over the Cartesian $x,y,z$ spin directions. 

As the experiment cannot access to the $(00l)$ scattering direction, to perform the integration over the full Brillouin zone we refer to the results of Ref. \cite{duan_incommensurate_2020} which suggest no dispersion along the $(00l)$ direction and intensity with a Gaussian distribution centered at $l=0$, with a width of $\sigma_{FWHM}\approx 0.5$ rlu. From this, we find an overall fluctuating moment of \mueff, which is consistent with what one would expect from the paramagnetic susceptibility of $3.6~\mu_B$~\cite{ikeda_single_2006} or of free-ion U$^{3+}$ of 3.62 $\mu_B$. By comparing the zero field and 11 T integrated Bragg peak intensities at the (110), (1$\bar{1}$0), and (020) Bragg peaks we find an upper limit of the field-induced ordered moment of $\mu=0.0(2)$ $\mu_B$. This is slightly at odds with what one would expect from magnetization, which for fields along the $b$ axis finds an induced moment on the order of $\mu=0.15$ $\mu_B$~\cite{ran_nearly_2019} at 11 T. A more detailed diffraction study using a large single crystal rather than a mosaic of coaligned samples would be more appropriate to resolve this moment. 

The lowest energy part of the scattering emerges from the $Y$ and $T$ points. The $\bm{Q}$-dependent intensity of the scattering can be directly calculated by assuming a form of $S(\bm{Q},\omega)$, where all spins in the unit cell have the same preferential spin orientation. The scattering takes the form
\begin{equation}
    S(\bm{Q})={\cal C}\sum_{\bm{\tau}}A_{Y,T}~\exp[-(\frac{(h-\tau_h)^2}{2\sigma_h^2} + \frac{(k-\tau_k)^2}{2\sigma_k^2}].
\end{equation}

The sum is over all of the $Y$ and $T$ points, where the intensity is assumed to have a Gaussian line shape of amplitude $A_T$ for $T$-points and $A_Y$ for $Y$-points. The peaks are of width $\sigma_h$ and $\sigma_k$ in the $h$ and $k$ directions, respectively. The prefactor ${\cal C}$ is defined by
\begin{equation}
    {\cal C}=\frac{g}{2} |F(\bm{Q})|^2 r_0^2 (1-\frac{\bm{Q}\cdot\hat{M}}{|Q|}).
\end{equation} 

Here, the preferential spin orientation $\hat{M}$ is constrained to be in the $(hk0)$ plane, with the tilt away from the $a$ axis towards the $b$ axis being denoted as $\theta_a$. No significant difference could be found between constraining the moment to the $(hk0)$ plane and an entirely free orientation. Fitting the cuts in Fig.~\ref{fig:cutfig}(c) and (d) simultaneously finds best fit values of $\theta_a$=24(3) degrees as depicted in Fig.~\ref{fig:cutfig}(e). The minimum at 24(3) degrees is shallow, but the scattering is clearly incompatible with a spin orientation beyond 30 degrees. This result unambiguously demonstrates that the spins contributing to the scattering  prefer to orient along the crystalline $\hat{a}$-axis within the \textit{ab} plane, despite no evidence of magnetic order, which is consistent with previous neutron studies~\cite{knafo_low-dimensional_2021} and bulk magnetometry indicating that the $\hat{a}$ is the magnetic easy axis~\cite{ran_nearly_2019}.

The character of the scattering in Fig.~\ref{fig:constE}(a) gives important information about the electronic band structure near the Fermi level $\epsilon_F$. As shown by the fit in~\ref{fig:cutfig}(c,d), the U$^{3+/4+}$ form factor is sufficient to describe the intensity modulation of the scattering with increasing $|Q|$, indicating that the spin density of electrons in the conducting band originates from the 5$f$ orbital. This is consistent with recent RIXS reports~\cite{christovam_stabilization_2024} suggesting a $5f^2$ configuration. Secondly, we note that the scattering is found at $Y=(0,0.6,0)$ and $T=(\frac{1}{2}\frac{1}{2}0)$ type points, but is absent at the $M=(\frac{1}{2}00)$ type points. This cannot be explained exclusively through the effect of spin polarization, as there are many symmetry-equivalent $M-$type points available in the $(hk0)$ scattering plane where no scattering is observed.

The scattering observed in this experiment confirms results from previous works~\cite{duan_incommensurate_2020,duan_resonance_2021,butch_symmetry_2022,raymond_feedback_2021,knafo_low-dimensional_2021}, and has direct physical consequences. The first is that at temperatures below T$_K \approx 40$ K, a dispersive magnetic excitation mode emerges which is peaked in intensity at 3.5(2) meV and is broad in energy ($\Delta E_{FWHM}\approx 1.5$ meV). While the scattering intensity may persist to zero energy transfer, at the lowest accessible energies the intensity becomes vanishingly small and the maxima of all excitations bands are fully gapped. There is no evidence for long-range magnetic order, but the fluctuating moment has preferential spin orientation along the $\hat{a}$ axis. The excitations themselves are constrained to the BZ edge, with no significant scattering at the $\Gamma$ point at all measured energy transfers. Finally, the momentum dependence of the scattering is described well by ionic U$^{3+/4+}$ and a total moment sum rule suggests that the scattering captures the majority of the relevant magnetic excitations. Though these results are informative, the interpretation of the origin of the magnetic scattering is less straightforward.

Previous INS studies of UTe$_2$ have characterized these excitations at non-integer $Q$ as antiferromagnetic spin fluctuations, suggesting a picture with dominant antiferromagnetic interactions between nearest neighbor U ions~\cite{duan_incommensurate_2020,duan_resonance_2021}. Subsequent analysis of neutron scattering spectra suggested the presence of both ferro and antiferromagnetic interactions in a spin-ladder model, where spin exchange interactions are ferromagnetic between the U$^{3+}$ chains along the $\hat{a}$-axis, ferromagnetic within the U-dimers along the $\hat{c}$-axis, and antiferromagnetic between ladders along the $\hat{b}$-axis~\cite{knafo_low-dimensional_2021}. Additionally, a pressure-induced antiferromagnetic ordered phase in UTe$_2$ was found at an experimentally modest hydrostatic pressure of about P$_{c}$=1.5 GPa~\cite{Thomas_Santos_Christensen_Asaba_Ronning_Thompson_Bauer_Fernandes_Fabbris_Rosa_2020,Knafo_2023_pressure}.  These studies report an incommensurate AFM ordering wavevector of $\bm{k} = (0.07,0.33,1)$~\cite{Knafo_2023_pressure}, and the close proximity of the ground state to an AFM ordered phase was taken as evidence for strong AFM magnetic interactions. This contradicts initial suggestions that spin-triplet superconductivity in UTe$_2$ is promoted by proximity to a ferromagnetic critical point. While spin-triplet superconductivity has been suggested to be possible in the case of AFM interactions~\cite{Tei_Mizushima_Fujimoto_2024}, the precise nature of the spin interactions is still an important open question. 

A magnon-like picture where the dispersion of the magnetic excitations originates from AFM interactions between spins raises a number of questions. The first is how magnons can emerge from a ground state that is paramagnetic. In this scenario, the excitations would be classified as paramagnons, where spins are proximate to local magnetic order due to strong spin correlations. Such a scenario is compatible with the broad signal observed in UTe$_2$. However, our data show that the magnetic response has no dependence on the applied magnetic field up to a relatively large field scale of $\mu_0 H$=11 T, contrary to what one would expect in the case of spins coupled with a Heisenberg interaction on the scale of 4 meV. Additionally, in the generic Heisenberg picture using linear spin wave theory the magnon dispersion minima occur at the magnetic ordering wave vector $\bm{k}_m$. Here, and in previous works~\cite{duan_incommensurate_2020,duan_resonance_2021,butch_symmetry_2022}, these minima are observed at the in equivalent $Y$ and $T$ points, requiring the further complexity of two different species of magnetic order or a multi-k structure. Because such a scenario quickly becomes quite complicated, we consider an alternative picture.

Neutron scattering from localized U-$f$ spins in heavy fermion compounds has previously been treated in the context of an extended Anderson lattice model~\cite{Brandow_1986,Brandow_1988}. In the case of UTe$_2$, this would mean that the Kondo effect hybridizes the U-$6d$, U-$5f$, and Te-$5p$ electronic bands near the chemical potential. In the extended Anderson lattice picture, the newly formed hybridized bands are constrained by symmetry to have saddle points with a significant density of states at high symmetry points at the Brillouin zone edge. This treatment provides an alternative description of the scattering, which is that of interband electronic transitions between regions with high density of state between the hybridized bands, which are at the zone edge and zone boundary~\cite{Auerbach_Kim_Levin_Norman_1988} and naturally account for the field-independence of the excitations. 

\begin{figure}
    \centering
    \includegraphics[width=1.0\columnwidth]{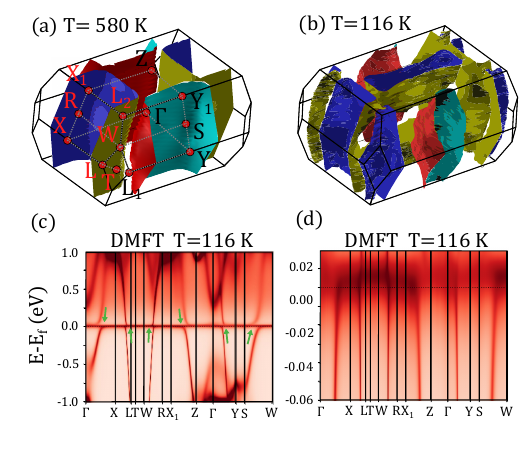}
    \caption{Fermi surfaces calculated from DFT+DMFT as described in the text at high temperature $T$=580 K (a) and low temperature $T$=116 K (b). The high temperature calculation in (a) The calculated low temperature electronic disperison throughout the full Brillouin zone is shown in (c), with the low energy scale most relevant to the quasiparticle excitations shown in (d). The green arrows in (c) highlight the anticrossings in the quasiparticle band structure that originate from $f$-band correlations.}
    \label{fig:band_structure}
\end{figure}

To explore this scenario, we compare to high resolution DFT+DMFT calculations of the band structure of UTe$_2$. These calculations were performed using the Rutgers code implemented in the WIEN2k package using the experimentally determined crystal structure~\cite{Hutanu_2020}, at temperatures of $T$=116 K, $T$=232 K, and $T$=580 K. A computational limitation of DFT+DMFT is the lower temperature limit, and thus the energy resolution of the band structure on the $\approx$10~meV scale. Thus, the detailed band structure near the chemical potential $\epsilon_f$ is not resolved in these calculations. Fig.~\ref{fig:band_structure}(b,d) plot the calculated quasiparticle band structure, which closely resembles our previously reported results in Ref.~\cite{mekonen_optical_2022}. We also note that these calculations are similar to those presented in Ref.~\cite{duan_incommensurate_2020}, with the primary difference being that there is no infinite Hubbard $U$ approximation. These calculations give insight into the hybridization associated with the $f-$band correlations, and while the precise energies within 10 meV of $\epsilon_f$ are unreliable due to the temperature constraint, the development of anticrossings of bands is evident. The anticrossings are denoted using arrows in Fig.~\ref{fig:band_structure}(c), and result in regions of high $f$-density of states at high symmetry points, between which interband excitations may occurr.

The calculated high temperature Fermi surface is shown in Fig.~\ref{fig:band_structure}(a), which finds two elliptical pockets similar to previous experimental and theoretical reports~\cite{eaton2024quasi,aoki2022first,Miao_2020}. At low temperature, Fig.~\ref{fig:band_structure}(b), the effect of hybridization may be observed with the shape of the pockets changing dramatically and a new pocket emerging around the $Z$-point. This can be compared directly to photoemission experiments performed at $T$=20 K, which report a heavy electron pocket at the $Z$ point~\cite{Miao_2020}, and quantum oscillation measurements which suggest both 2D and 3D Fermi surface character, with conflicting conclusions as to the presence of such a pocket~\cite{Broyles_Ran_2023,Weinberger_Eaton_2024}. In both plots, the blue/red surfaces originate from Te-$5p$ band, and the blue/yellow surfaces originate from the U-$6d$ band. 

\begin{figure}
    \centering
    \includegraphics[width=1.0\columnwidth]{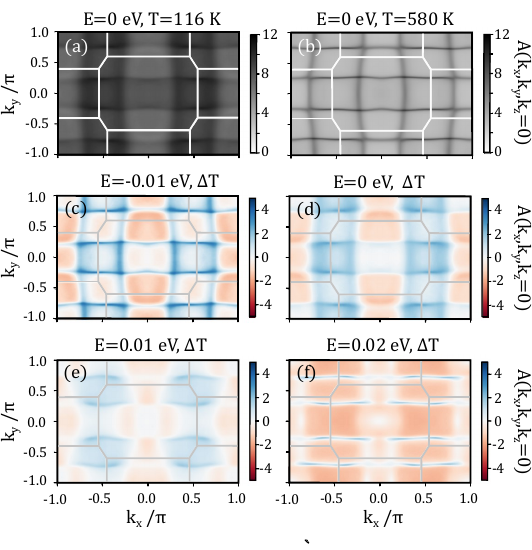}
    \caption{Constant energy $k_z$=0 slices of the spectral function $A(\bm{k},E)$, with energy difference from $E_f$ being denoted in the title of each subplot. Directly calculated Fermi surfaces plotted for $T$=116 K (a) and $T$=580 K (b) show the effect of hybridization and the accumulation of $f$-weight at low temperature. The difference between the two temperatures is shown at $E-E_f$=-10 meV (a), 0 meV (b), 10 meV (c), and 20 meV (d), which is taken as a measure of the distribution of $f$-electron character.}
    \label{fig:dmft_slices}
\end{figure}

To examine the redistribution of $f-$weight at low temperature, we examine constant energy $k_z$=0 slices in Fig.~\ref{fig:dmft_slices}. Figs.~\ref{fig:dmft_slices}(a) and (b) show the $k_z$=0 Fermi surface for $T$=116 K and $T$=580 K respectively. We take the temperature difference of the spectral function $A(\bm{Q},\omega)$ to be a measure of the U-$5f$ band weight, which accumulates in pockets centered about the $X$ point at E=E$_f$, as shown as a function of energy transfer in Figs.~\ref{fig:dmft_slices}(c-f). The $f$-weight disperses towards the zone corner at higher energy transfers, which is suggestive of the origin of the dispersion of the excitations observed in INS. This dispersion, which originates from hybridization, is also anisotropic in the $(hk0)$ plane, as most clearly demonstrated at the chemical potential in Fig.~\ref{fig:dmft_slices}(d).

Using the distribution of the $f$-electron character presented in Fig.~\ref{fig:dmft_slices}(d), it is tempting to assign potential nesting vectors between regions within high spectral weight. This would lend support to the idea of an emergent charge density wave which has been observed in scanning tunneling microscopy (STM) studies, but has not been observed in bulk scattering measurements~\cite{aishwarya2023magnetic,aishwarya2024melting,kengle2024absence,Kengle_xray_2024,theuss2024absence}. As no such wavevector is observed in the elastic component of the magnetic scattering, such a phase would need to be a surface state rather than in the bulk.

Due to the energy resolution of the calculation, it is not possible to determine the precise scattering pathways that would lead to the measured $S(\bm{Q},\hbar\omega)$ in INS. However, the calculations do show that the hybridized band structure features parabolic-like bands with extrema at the zone boundary, as suggested in the Anderson picture. Two essential conditions are required for interband scattering, the first is the presence of a high density of states near the chemical potential at the high symmetry points at which scattering is observed, and the second being anisotropy in the band hybridization resulting in the lack of scattering at the $X$ point, both of which are captured in the DFT+DMFT calculations. 

Our work resolves the magnetic excitations in UTe$_2$ that peak at $\hbar\omega$=3.5(2) meV, which we propose originate from electronic excitations near or across the Kondo hybridization gap. This conclusion is also supported by our previous work~\cite{butch_symmetry_2022}, in which the temperature dependence of the magnetic scatting intensity is linked to the Kondo coherence tmperature. Here, the excitations capture a total fluctuating magnetic moment of \mueff, which is approximately compatible with what one would expect for free-ion U$^{3+}$. Our analysis shows that their origin is not from magnetic interactions, but no conclusion is made regarding the proximity of UTe$_2$ to any particular ordered state, or the nature of its magnetic interactions which have been reported as both ferromagnetic~\cite{ran_nearly_2019}, and antiferromagnetic~\cite{duan_incommensurate_2020}. This is consistent with previous DFT+U studies, which suggest that ferromagnetic inter dimer interactions between U-ions may stabilize spin-triplet superconductivity~\cite{Xu_Sheng_Yang_2019,Shishidou_Suh_Brydon_Weinert_Agterberg_2021}. The scattering captures the full expected magnetic moment in the system, and we expect that this scenario could be supported by further DMFT and calculations of $\chi''(\bm{Q},\omega)$ similarly to the studies of Refs.~\cite{fuhrman_interaction_2015,goremychkin_coherent_2018}. The DMFT results can be verified directly through photoemission high resolution experiments at low temperatures, which would be able to resolve the elliptical pockets predicted in our calculations. Additionally, we hope that this work will motivate the further development of DFT+DMFT methods to enable calculations of $\chi''(\bm{Q},\hbar\omega)$ in the meV scale energy regime, which is relevant not only in the case of UTe$_2$ but many other heavy fermion systems with Kondo hybridization. 

\subsection{Methods}
The INS experiment was performed on a coaligned mosaic of 82 crystals of UTe$_2$ with a total mass of 1.1 g. Most crystals were grown by chemical vapor transport~\cite{yao_controllable_2022}, with eight of total mass $\approx80~$mg being grown by a salt flux method~\cite{aoki_molten_2024,sakai_single_2022}. All batches of crystals were screened by either transport measurements or susceptibility, with a range of critical temperatures of $T_c\in\{1.8,2.0\}$ K indicating high sample quality. The crystals were mounted on a oxygen-free copper sample holder using CYTOP (AGC Chemicals Company) and aligned using a x-ray Laue diffractometer.

The INS measurement was performed using the CAMEA spectrometer~\cite{lass_commissioning_2023} at the Paul Scherrer Institut to measure the inelastic scattering in an energy transfer range of $\hbar\omega\in\{-1,6\}$ meV. The sample environment was the MB11 $\mu_0 H$=11 T vertical cryomagnet with a dilution refrigerator insert with a base temperature of $T$=45 mK throughout the experiment based on the sample thermometer. All measurements were performed in the $(hk0)$ scattering plane, with the magnetic field  oriented along the $\hat{c}$ direction. For the zero-field configuration, the total counting time was 61 hrs, and for the 11 T configuration 64 hrs. The scattering was normalized to absolute units of (b/meV/sr/U) using a vanadium standard of known mass. All analysis was performed using the MJOLNIR software~\cite{lass_mjolnir_2020,Lass_MJOLNIRPackage_abehersan_2024}. 

\section{Author Contributions}
Syntheis of samples used in this study was performed by C.F., P.C., and G.S. The INS experiment was conceived by N.B. and M.J., and performed by T.H., P.C., and J.A.R., with assistance from D.G.M. and J.L. DFT+DMFT calculations were performed by C.J.K. and G.K. The mauscript was written by T.H. and N.B. with input from all authors. 

\section{Competing Interests}
The authors declare no competing interests. 

\section{Data Availability}
All data are available upon reasonable request to Thomas Halloran.

\section{Acknowledgements}

This work is based on experiments performed at the Swiss spallation neutron source SINQ (proposal number 20230495), Paul Scherrer Institut, Villigen, Switzerland. Work at UMD was supported by NIST and by the National Science Foundation under the Division of Materials Research Grant NSF-DMR 2105191. MJ acknowledges funding by the Swiss National Science Foundation through the project "Berry-Phase Tuning in Heavy f-Electron Metals (\#200650). Identification of commercial equipment does not imply recommendation or endorsement by NIST. GK was supported by DOE BES CMS program. C.-J. K. was supported by the National Research Foundation of Korea (NRF) (Grant No. NRF-2022R1C1C1008200). 
\bibliography{refs}

\end{document}